\documentclass[aoas]{imsart}
\usepackage{amsmath}
\usepackage{textcmds}
\usepackage{amssymb}
\usepackage{amsthm}
\usepackage{graphicx}
\usepackage[]{xcolor}
\usepackage{subcaption}
\RequirePackage[authoryear,square]{natbib}
\RequirePackage[colorlinks,citecolor=blue,urlcolor=blue]{hyperref}

\begin{document}
\begin{frontmatter}

\title{Plans for acceptance sampling by attributes when observations are destructive}

\runtitle{Acceptance sampling plans for destructive testing}
\begin{aug}
\author[A,B]{Hugalf Bernburg\orcid{0009-0002-0954-092X}}
\and
\author[A]{Katy Klauenberg}
\address[A]{Physikalisch-Technische Bundesanstalt, Abbestr. 2-12, 10587 Berlin, Germany}
\address[B]{ Friedrich Schiller University Jena,
Institute for Mathematics,
Ernst-Abbe-Platz 2,
07743 Jena, Germany}
\date{\today}
\maketitle
\end{aug}
\begin{abstract} 
The international standard ISO 2859-2 provides plans for acceptance sampling by attributes, that ensure a defined quality level in isolated lots using the hypergeometric distribution. In destructive testing, the sample itself is damaged or changed such that the quality of an entire lot is less relevant than the quality of the lot that remains after removing the sample. Examples include assessing the germination of seeds and the conformity of in-service utility meters.

This research highlights that the hypergeometric distribution cannot describe the frequentist consumer’s risk of accepting a remaining lot with unsatisfactory quality. Consequently, sampling plans as those provided in ISO 2859-2 are ill-suited to assess the remaining lot when sampling destructively. In contrast, Bayesian statistics inherently infers the lot’s quality after sampling. Using a reference prior, we show that sampling plans provided by ISO 2859-2 result in high specific consumer’s risk for small remaining lots.

The ISO 2859-2 being ill-suited, we design plans for destructive sampling that limit the (Bayesian) specific consumer’s risk. To tabulate these plans in a similar way to ISO 2859-2, we propose a new representation that fixes the remaining lot size $N-n$ rather than the sample size $n$. This generalizable, concise and efficient representation is suitable for future standardization of destructive sampling.
\end{abstract} 


\end{frontmatter}
\section{Introduction}\label{sec:intro} 

Acceptance sampling is a means to limit the risk of accepting lots with unsatisfactory quality \citep{ISO28590}.
Acceptance sampling is pervasive and adopted, for example, as quality control measure when monitoring manufacturing processes, or as quality-based decision to accept or reject purchased goods in international trade \citep{Uhlig24}.
It is also used by governmental food safety agencies to ensure that contaminant levels are below legal limits \citep{Uhlig24} or by conformity assessment bodies to ensure that measuring instruments conform with requirements in legal metrology.
 
 For acceptance sampling, ISO TC 69/SC 5 administrates 28 standards.
 Among these, the ISO 2859 series on attribute sampling \citep[clause 4.2.12]{ISO3534-2} provides tabulated sampling plans fulfilling some frequentist risk criteria.
 The ISO 2859 series is applied, for example, to control the quality 
 of industrial production processes, 
 of medical devices, 
 or of regulated measuring instruments.
Examples of regulated instruments are utility meters to be placed on the EU market \citep{WelmecGuide8.10} and kept under surveillance \citep{OIMLG20}.
 For lots in isolation, i.e.\ lots which are separated from other lots, the standard ISO 2859-2 \citep{ISO2859-2} shall limit the consumer's risk, that is, it shall limit the probability of accepting lots with limiting quality (LQ), cf.\ \citep[clause 4.6.13]{ISO3534-2} and \citep[clause 3.1.1]{ISO2859-2}.
  
Destructive testing is often used in material science and structural engineering, where it is considered to be \q{the process of inspecting, evaluating and measuring the properties of materials or systems in a manner which can change, damage or destroy the properties or affect the service life of the test specimen} \citep[p.\ 9]{Wilson84}.
 Examples are hardness tests, crash tests, tests for explosives.
 Destructive testing is applied accordingly to measurement systems \citep{Gorman02} and transferred 
 to students taking tests in educational research \citep{Avellaneda24}, to psychology \citep{Anderson96}, 
 and to seed sciences \citep{herman2013Hypergeometric}. 
Destructive testing is applied when no alternative exists, or when it is more reliable, more cost efficient, simpler or otherwise superior to non-destructive testing.

This research focuses on acceptance sampling for destructive testing, which commonly occurs in the assembly and manufacturing industry \citep{Delgadillo07} and when in-service utility meters or seed lots are to comply with regulatory requirements \citep{OIMLG20,herman2013Hypergeometric}.

Clearly, destructive sampling  is an instance of sampling without replacement.
 When sampling without replacement and counting the nonconforming items in that sample, the usual quantity of interest is the quality of the whole lot, which is given as percentage of nonconforming items.
 Given this quality, the number of nonconforming items in a random sample is well-known to follow the hypergeometric distribution \citep[chapter 6]{Johnson92}.
 This distribution is also the basis for the sampling plans given in ISO 2859-2 for nonconforming items.
 However, when sampling is destructive, the quality of the whole lot is of little interest.
 The quantity of interest is rather the quality of the remaining lot, i.e.\ the percentage of nonconforming items in the lot downsized by the sample. 

 Destructive sampling has been addressed in the literature. For series of lots, skip-lot sampling is suggested (see e.g.\ \cite{Hsu77}, \citep{ISO2859-3}, and for a Bayesian approach \cite{Phelps82}).  
 However, related research did not consider the quality of the remaining lot.
 Also \cite{Hald60} includes destructive sampling, but sampling plans are designed to optimize costs and thus consider the expected quality of the remaining lot. \cite{wallenius1967sampling} does consider the remaining lot, but limits its absolute number of nonconforming items irrespective of the size of the remaining lot. Consequently, \cite{wallenius1967sampling} does not limit the percentage of nonconforming items of the remaining lot.

 Currently, no standard equivalent to ISO 2859-2 exists that provides sampling plans for acceptance sampling by attributes, when the quality of the remaining lot is of interest, as is the case for destructive sampling.
 The sampling plans in ISO 2859-2 themselves are ill-suited  to assess the remaining lot when sampling destructively, which we highlight in Section~\ref{SecISO}.
 They provide insufficient consumer protection.
 In fact, the hypergeometric distribution cannot quantify the frequentist consumer's risk about the remaining lot.

 The Bayesian approach on the other hand is well-suited for destructive sampling because a posterior distribution can naturally describe the quality in the remaining lot and the corresponding sampling distribution is known.
 Within the Bayesian approach, there are ample possibilities to design sampling plans suitable for destructive sampling, which will be summarized briefly in Section~\ref{SecBayes}. 
 One of these possibilities to design sampling plans is to limit the Bayesian risk known as the specific consumer’s risk \citep{JCGM106}. 
 It is described in Section~\ref{SecBayes},
 where the resulting sampling plans are also given and compared to those provided in ISO 2859-2.

Finally, in Section~\ref{SecTabs} a representation is suggested to tabulate sampling plans for destructive sampling.
 Such tabulated plans provide a balance between efficient sampling and concise presentation, they are simple to use and suitable for standardization.

\section{Inappropriateness of using sampling plans as those in ISO 2859-2 for destructive sampling}\label{SecISO}

The standard ISO 2859-2 designs acceptance sampling plans for inspection by attributes and for lots in isolation.
The standard allows for two cases: firstly, inspection for nonconforming items, which is considered in this article and secondly, inspection for nonconformities, which is not considered here.
When one inspects for nonconforming items, the sampling plans are defined by a sample size $n$, an acceptance number Ac, and the following operational instructions:
Take a random sample of size $n$ without replacement from the lot and count the nonconforming items. 
If you observe less than or equal to Ac nonconforming items, accept the lot; otherwise reject it \citep[annex A]{ISO2859-2}.

\begin{figure}
	 \includegraphics[width=.85\linewidth]{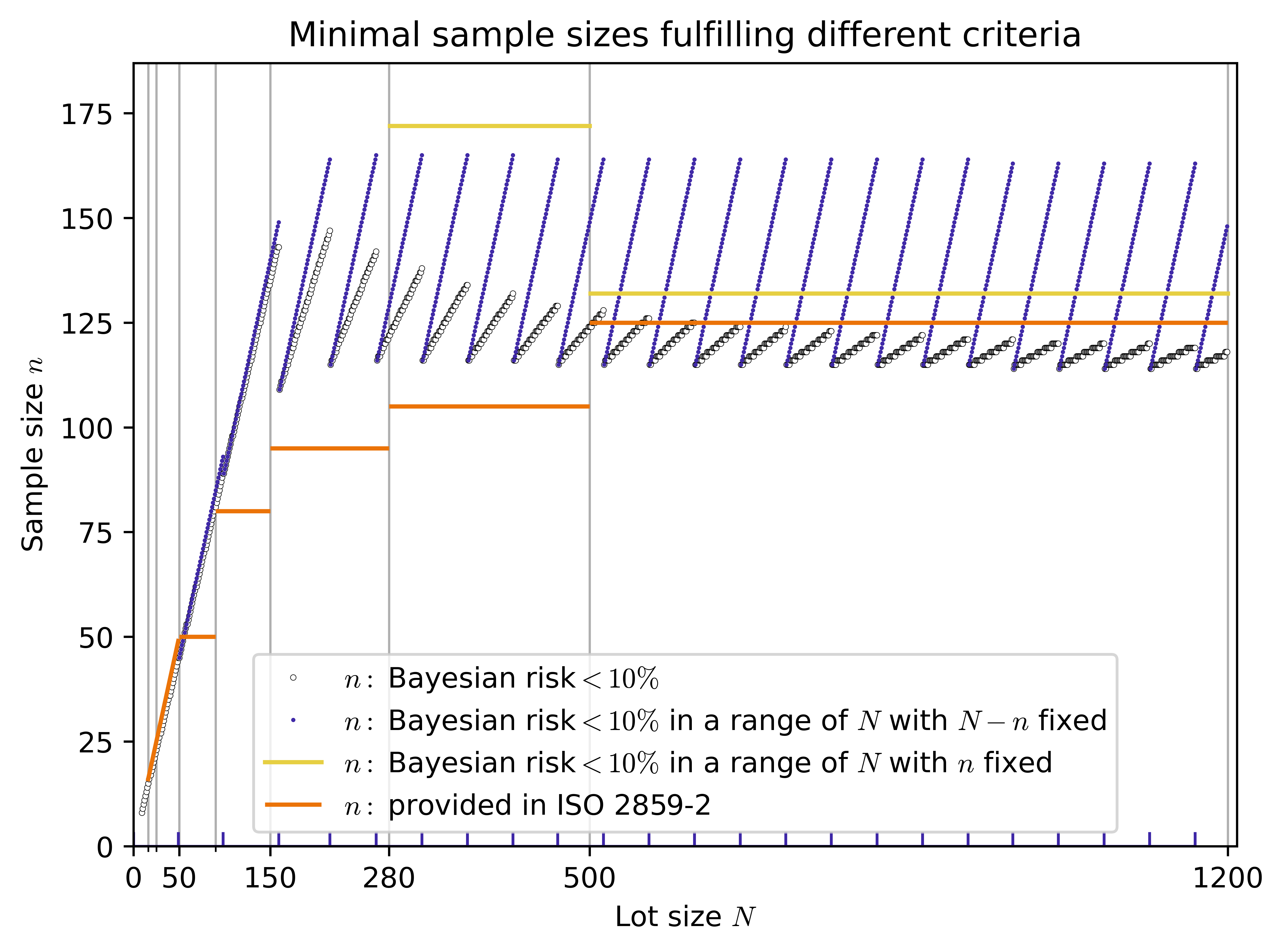}
 \caption{A limiting quality of LQ = 2~\% and acceptance number $\text{Ac}=0$ is assumed.
 Displayed in orange are the sample sizes provided in the standard ISO 2859-2 for ranges of whole lot sizes.
 The sample sizes limiting the risk \eqref{EqBayesRisk} by 10~\% for each whole lot size are displayed in black. Those limiting it for a range of lot sizes simultaneously are displayed in yellow. 
The ranges are the same as used in the standard ISO 2859-2.
The sample sizes limiting the risk \eqref{EqBayesRisk} by 10~\% such that $N-n$ is fixed for a range of lot sizes is displayed in blue. 
The ranges for the latter are chosen such that the difference between neighboring optimal sample sizes (black) is small in each range. 
}
\label{FigOptimSP}
\end{figure}

For simplicity, we call the pair ($n$, Ac) a sampling plan and omit repeating the operational instructions.
Some plans from the standard are displayed in orange in Figure~\ref{FigOptimSP} for LQ = 2~\%.
Each sampling plan is designed for a specific range of lot sizes 
and LQs (see \citep[tables 8, 9, annex B and section 4.2]{ISO2859-2} for details).
It is designed such that the consumer's risk is limited to approximately 10~\% \citep[clause 3.1.1]{ISO2859-2}.
The consumer's risk is the probability of accepting a lot with limiting quality LQ \citep[clause 3.1.2 note 1]{ISO2859-2}.
When $Y$ denotes the number of nonconforming items in a random sample and $k_{whole}$ denotes the number of nonconforming items in the whole lot, the consumer's risk can be written as
\begin{align}
    P(Y\leq \text{Ac} \mid k_{whole}=\lceil \text{LQ}\cdot N\rceil,N,n) \stackrel !\leq 10 ~\%,\label{EqFreqRisk}
\end{align}
with the ceiling function $\lceil \cdot\rceil$ rounding its arguments up to the next integer. 
This probability is also known as type I error in statistical hypothesis testing which is defined as the maximal probability of erroneously rejecting the null hypothesis
\begin{align}\label{eq:H0}
    H_0\!:k_{whole}\geq \lceil \text{LQ}\cdot N\rceil.
\end{align}  
At the same time, the plans of the standard are designed such that the producer's risk (type II error) is limited to 5~\% \citep[clause 3.1.4, clause R3 in annex B]{ISO2859-2}.

To calculate the consumer's and producer's risks given in tables 8 and 9 of the standard ISO 2859-2, the hypergeometric distribution is used \citep[annex A]{ISO2859-2}:
\begin{equation}\label{EqHG}
  P(Y=y|k_{whole},N,n)=\text{Hypergeometric} (y;N,k_{whole},n)= \frac{\binom{k_{whole}}{y}\binom{N-k_{whole}}{n-y}}{\binom{N}{n}}.
 \end{equation}

We emphasize that the standard ISO 2859-2 designs sampling plans assuring low acceptance probability whenever the quality of the  \textit{whole} lot --- a lot that includes the sample \citep[annex~A]{ISO2859-2} --- is unsatisfactory, i.e.\ whenever the quality is worse or equal to LQ.
It does not design sampling plans assuring low acceptance probability when the quality of a \textit{remaining} lot --- a lot reduced by the sample --- is unsatisfactory.
However, when sampling is destructive, the quantity of interest may rather be the quality of the remaining lot.
To assess this quality, sampling plans are needed, and existing sampling plans do not limit the consumer’s risk of an unsatisfactory quality in the remaining lot.

\begin{figure}
\centering
	\includegraphics[scale=.8]{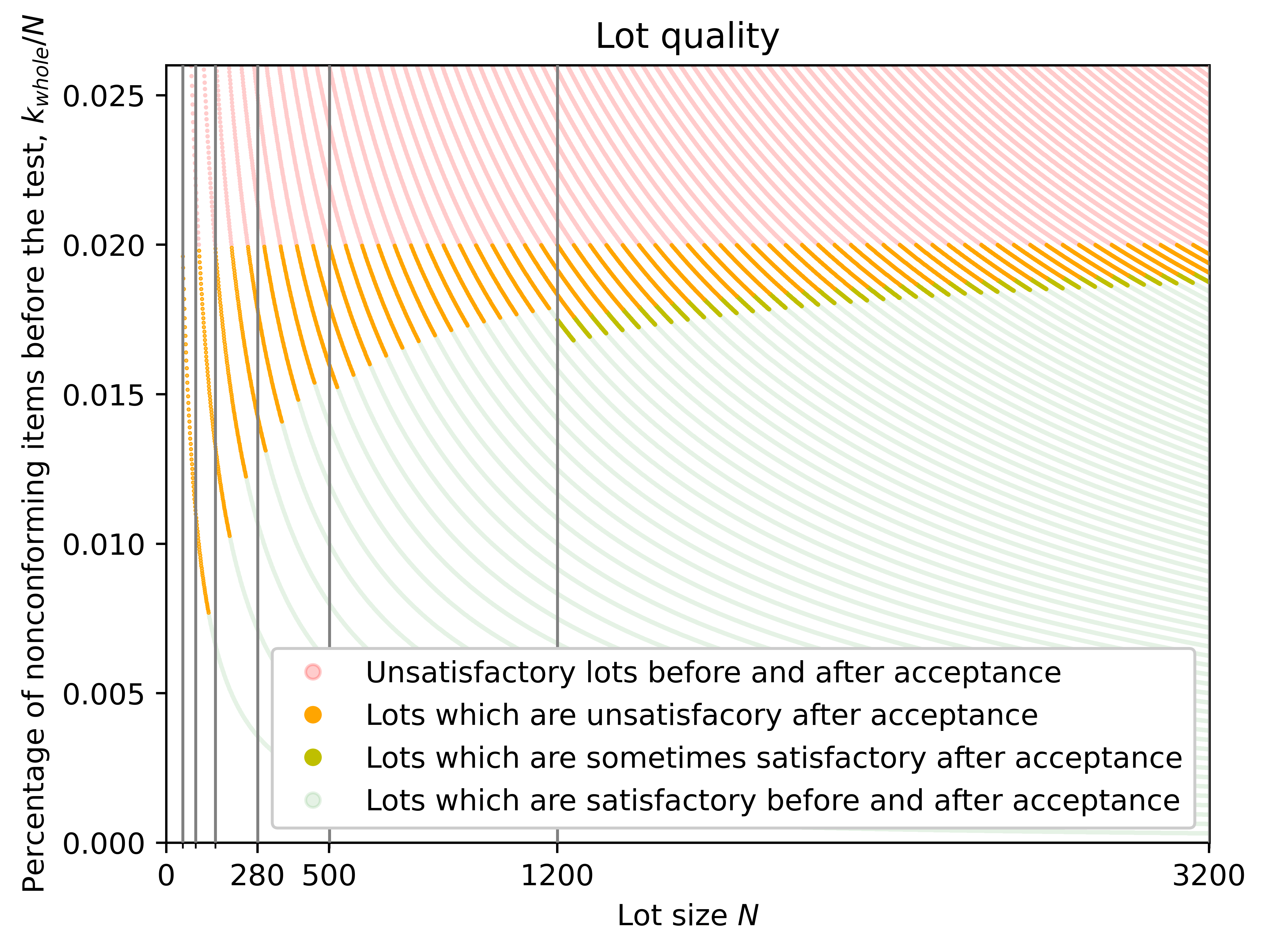}
	\caption{Displayed are lots of different quality and size. 
    Every point is one lot with a specific number of nonconforming items.
    The color indicates the quality of the lot before and after taking an acceptable sample according to ISO 2859-2 table~9 with LQ $=2~\%$.
    Displayed in red are lots whose quality is unsatisfactory before and after taking such sample.
    Orange are those lots whose quality is no longer satisfactory after acceptance ($k_{rem}\geq\lceil 0.02\cdot(N-n)\rceil$).
    Olive green are those lots which can be either satisfactory or not after acceptance depending on how many nonconforming items are in the sample.
    Displayed in green are  lots for which the quality is satisfactory before and after taking a sample ($k_{rem}<\lceil 0.02\cdot(N-n)\rceil$).
    }
	\label{FigQualLevel}
\end{figure}

Let $k_{rem}$ denote the number of nonconforming items in the remaining lot.
The following states that the consumer’s risk of an unsatisfactory quality in the remaining lot shall be limited by 10~\%:
\begin{equation}\label{EqFreqRiskDestr}
  P(Y\leq \text{Ac}\mid k_{rem}=\lceil \text{LQ}\cdot (N-n)\rceil,N,n)\stackrel!\leq 10~\%.
\end{equation}
Lots which have satisfactory quality before sampling ($k_{whole}<\lceil \text{LQ}\cdot N\rceil$) may have unsatisfactory quality after sampling ($k_{rem}\geq\lceil \text{LQ}\cdot (N-n)\rceil$).
Figure~\ref{FigQualLevel} classifies the quality of lots before and after taking a sample for a range of lot sizes and for the sample sizes given in the standard ISO 2859-2. 
The green (red) points display lots with (un-)satisfactory quality before and after taking the sample. 
However, the orange points display lots with satisfactory quality before sampling but unsatisfactory quality after taking an acceptable sample.
The olive green points display lots with satisfactory or unsatisfactory quality after acceptance depending on the number of nonconforming items in the sample.
To calculate and limit the risk of accepting unsatisfactory remaining lots, the red and (contrary to ISO 2859-2) also the orange lots should count as unsatisfactory.
The olive green lots should not be considered as satisfactory either. 

We conclude that the percentage of nonconforming items in the whole lot needs to be (much) lower than the LQ so that acceptance of the lot results in a remaining lot of satisfactory quality ($k_{rem}< \text{LQ}\cdot \lceil (N-n)\rceil$).
The word \q{much} applies especially to small (remaining) lots. 
The ISO 2859-2 sampling plans are based on the quality before taking a sample and are thus not capable to assess the quality for remaining lots.
The authors recommend future editions to explicitly state this limitation as is done in other standards 
 or early publications (like \citep[clause 1]{ISO28594}  and \cite[pp. 10, 26]{Dodge1944Sampling}).
We are not aware of any standard that assesses the quality of the remaining lot. That is, there is no standard applicable for destructive sampling for which the remaining lot is of interest.
 
 More generally, the risk in equation \eqref{EqFreqRiskDestr} of an insufficient quality of the remaining  lot cannot be described by the hypergeometric cumulative distribution function.
 This is because $k_{rem}=k_{whole}-Y$, and hence the event on which one conditions on in equation~\eqref{EqFreqRiskDestr}, involves the random variable $Y$. 
\cite{herman2013Hypergeometric} observed already that \q{the hypergeometric distribution overestimates our confidence in low [numbers of nonconforming items] in the residual lot} and contrasted the probability~\eqref{EqHG} with $P(k_{rem}=1|k_{whole}=1)$ for illustrative examples. 

 To the best of the authors' knowledge there is no combinatorial description of the risk~\eqref{EqFreqRiskDestr}.
 If interpreted as prediction interval for samples from a binomial distribution, different inferences exist. 
 However, these lead to different interval boundaries (see \cite{patel1989prediction} for a review) and would thus lead to different sampling plans. 
 Also the interpretation itself, with $k_{rem}$ representing a previous sample, may be questionable. 
 The authors are not aware of other approaches to quantify the consumer's risk~\eqref{EqFreqRiskDestr}. 
 We thus abandon
 the aim to limit the consumer's risk \eqref{EqFreqRiskDestr}
 for destructive sampling and turn to designing sampling plans which limit Bayesian consumer's risks in Section~\ref{SecBayes}.

\section{The Bayesian approach to destructive sampling}\label{SecBayes}
In Bayesian sampling procedures one can select the sample size such that some utility is maximized (see \cite{Insua20} for a recent review of Bayesian decision making in reliability).
Other Bayesian approaches aim at limiting risks, for example the probability of a bad remaining lot given its acceptance.
A list of different risks for acceptance sampling by attributes is given in \cite[section 2]{Uhlig24}.
We focus on the specific consumer's risk
\begin{align}\label{EqBayesRisk}
    P( k_{rem}\geq \lceil \text{LQ}\cdot (N-n)\rceil\mid y,N,n),\quad y\leq\text{Ac},
\end{align}
which is defined more generally in the guide \citep{JCGM106} and applied, for example, in reliability or acceptance sampling (e.g.\ \cite{Wilson21,Uhlig24}). We calculate this risk based on the posterior distribution
\begin{align}\label{EqPost}
 P(k_{rem}\mid y,N,n)&=P(k_{whole}-y\mid y,N,n)\\\nonumber
 &\propto P(y\mid k_{whole},N,n)P(k_{whole}),
\end{align}
where $\propto$ denotes proportionality in the variable $k_{whole}$.
\sloppy
 The sampling distribution $P(y|k_{whole},N,n)$ is known to be the hypergeometric distribution~\eqref{EqHG} when we inspect $n$ items without replacement from a lot of size $N$ with $k_{whole}$ nonconforming items (see, e.g., \cite{Dyer93})\footnote{References like \cite{Dyer93} usually consider $k_{whole}$ only and do not distinguish $k_{rem}$ and $k_{whole}$ in equation~\eqref{EqPost}.}.
 The prior distribution $P(k_{whole})$ is chosen by the consumer when calculating the consumer's risk, and is chosen by the producer when calculating the producer's risk.
That is, the prior may depend on the decision maker.

Equation~\eqref{EqPost} reveals the suitability of the Bayesian approach for destructive sampling for the following two reasons:
Firstly, the sampling distribution conditions on the quality in the whole lot $k_{whole}$ and is therefore known through combinatorial methods --- thus avoiding the difficulties pointed out in Section~\ref{SecISO}.
Secondly and per construction, the posterior distribution encodes the beliefs after sampling and can thus easily target the quality in the remaining lot $k_{rem}$. 

Let us now compare Bayesian sampling plans that are designed for destructive sampling that assess remaining lots with those provided in the standard ISO 2859-2.
For this purpose, we design sampling plans for destructive sampling that limit the specific consumer's risk~\eqref{EqBayesRisk} by 10~\%
 when the so-called non-informative prior $P(k_{whole})=1/N$ is used.
This design criterion can be viewed as a Bayesian analogue to limiting the frequentist consumer's risk~\eqref{EqFreqRiskDestr}.
It is applied for example in reliability (e.g.\ \cite{Wilson21}) or acceptance sampling \cite{Uhlig24}.
The uniform prior is the reference prior  for the hypergeometric sampling distribution \citep[p.\ 28]{bernardo2003bayesStatis}.
The beta-binomial distributions contain the uniform prior and are the conjugate distributions for the hypergeometric sampling distribution \citep{Dyer93}.
Using a uniform prior is thus reasonable to make a fair comparison with frequentist sampling plans and is also convenient computationally\footnote{Since sampling plans are sensitive to the risk threshold, different numerical approximations of the risk may occasionally result in slightly different sampling plans.
To calculate the risk~\eqref{EqBayesRisk}, we use \q{stats.betabinom.cdf} from the SciPy package. }. 
However, the consumer may as well use more informative priors to generate sampling plans.

There are several sampling plans limiting the risk~\eqref{EqBayesRisk} by 10~\%, among which the producer may choose one.
This plan may be the one which maximizes the producer's utility, limits some risk or assurance \citep{Wilson21}, or it may even be a sequential plan. 
We will only consider zero failure sampling plans, i.e.\ plans with $\text{Ac}=0$.
Single sampling plans limiting the risk~\eqref{EqBayesRisk} by 10~\% with Ac $\geq 1$ require larger samples such that for small lots no items remain.
Figure~\ref{FigOptimSP} displays in black the minimal sample size such that the risk \eqref{EqBayesRisk} is limited by 10~\% for LQ~$=2~\%, \text{Ac}=0$ and for different initial (whole)  lot sizes. 
For the  lot sizes $54<N\leq 500$, the sample sizes provided by ISO 2859-2 are smaller than those limiting the Bayesian consumer's risk~\eqref{EqBayesRisk}. 
That is, for the plans provided in ISO 2859-2 the Bayesian consumer's risk~\eqref{EqBayesRisk} of bad quality in the remaining  lot will be larger than 10~\%. 
In fact, the risk peaks at 44~\% for the plan (50, 0) at $N=90$.
Similar results are expected for other values of LQ and small remaining  lot sizes.

We conclude that the Bayesian approach provides a simple method to design sampling plans for destructive sampling where the frequentist approach  \eqref{EqFreqRiskDestr} has difficulties.
The sampling plans provided in the standard ISO 2859-2 do not limit the specific consumer's risk of bad quality in the remaining  lot.
Despite recent efforts to include the Bayesian approach when standardizing acceptance sampling by attribute \citep{Uhlig24}, no standard is available yet.
In particular, no standard provides sampling plans for destructive sampling in isolated lots based on the Bayesian approach.

\section{Towards standardizing destructive sampling plans}\label{SecTabs}

 Since the currently available standards are ill-suited to assess the remaining lot when sampling destructively, the need arises to tabulate sampling plans for destructive testing, such as those suggested in Section~\ref{SecBayes}.
 While software is capable of providing the most efficient sampling plans, tabulated sampling plans have several advantages: 
 they provide an overview for two discretely varying input quantities (LQ and  lot size here), 
 they are simple to use,
 and they are immune to software security issues.
 Tabulated sampling plans prevail in standardization, which are in turn widely used (cf.\ Section~\ref{sec:intro}) and are important in regulated areas such as legal metrology and food safety.
Considering both, software and tables can  facilitate negotiations on the choice of the sampling plan.

\subsubsection*{Obstacles when tabulating sampling plans for destructive testing}
 Figure~\ref{FigOptimSP} displays sampling plans with Ac~$=0$ for LQ = 2~\% and for lots with sample sizes below 1200. 
 The black circles are the minimal sample sizes limiting the risk~\eqref{EqBayesRisk} by 10~\%.
 To concisely tabulate plans for destructive sampling, we select
 ranges of lot sizes and require one sample size to limit the risk~\eqref{EqBayesRisk} for this whole range. Such tabulated sample sizes are larger than the minimal ones, and
 are given in yellow in Figure~\ref{FigOptimSP}. 
 The ranges are the same as in the ISO 2859-2 \citep[table 9]{ISO2859-2}: 16-25, 26-50, 51-90, 91-150, 151-280, 281-500, 501-1200. 
We see that fixed sample sizes for a range of lot sizes are large when the lot is small.
In particular, they are much larger than the optimal plans when LQ = 2~\% and $N\leq500$.
See for example the sample size $n=172$, which for the acceptance number Ac~$=0$ limits the risk~\eqref{EqBayesRisk} by 10~\% for all $N\in\{281,282,...,500\}$ simultaneously.
Optimal sample sizes for each lot size from this range are below 138.
A more extreme example is the range $N \in \{151, 152, ..., 280\}$ for which no common sample size $n$  limits the risk~\eqref{EqBayesRisk} by 10~\% and leaves a remaining lot (i.e.\ there is no $N$ in that range such that $N>n$).
The same holds true for every listed range of lot sizes below 151.
 
 The sample sizes provided by the ISO 2859-2 that correspond to these ranges are displayed in the same Figure~\ref{FigOptimSP} in orange. These sample sizes,
 the optimal sample sizes for destructive sampling (black circles), 
and the sample sizes for destructive sampling fixed for a range of lots (yellow)
differ substantially for small lots and are similar when the lots are large.
We can further see that the sampling plans from the ISO 2859-2 are monotone increasing in $N$, while the plans for destructive sampling are not.

\begin{figure}
     \centering
     \includegraphics[width=.85\linewidth]{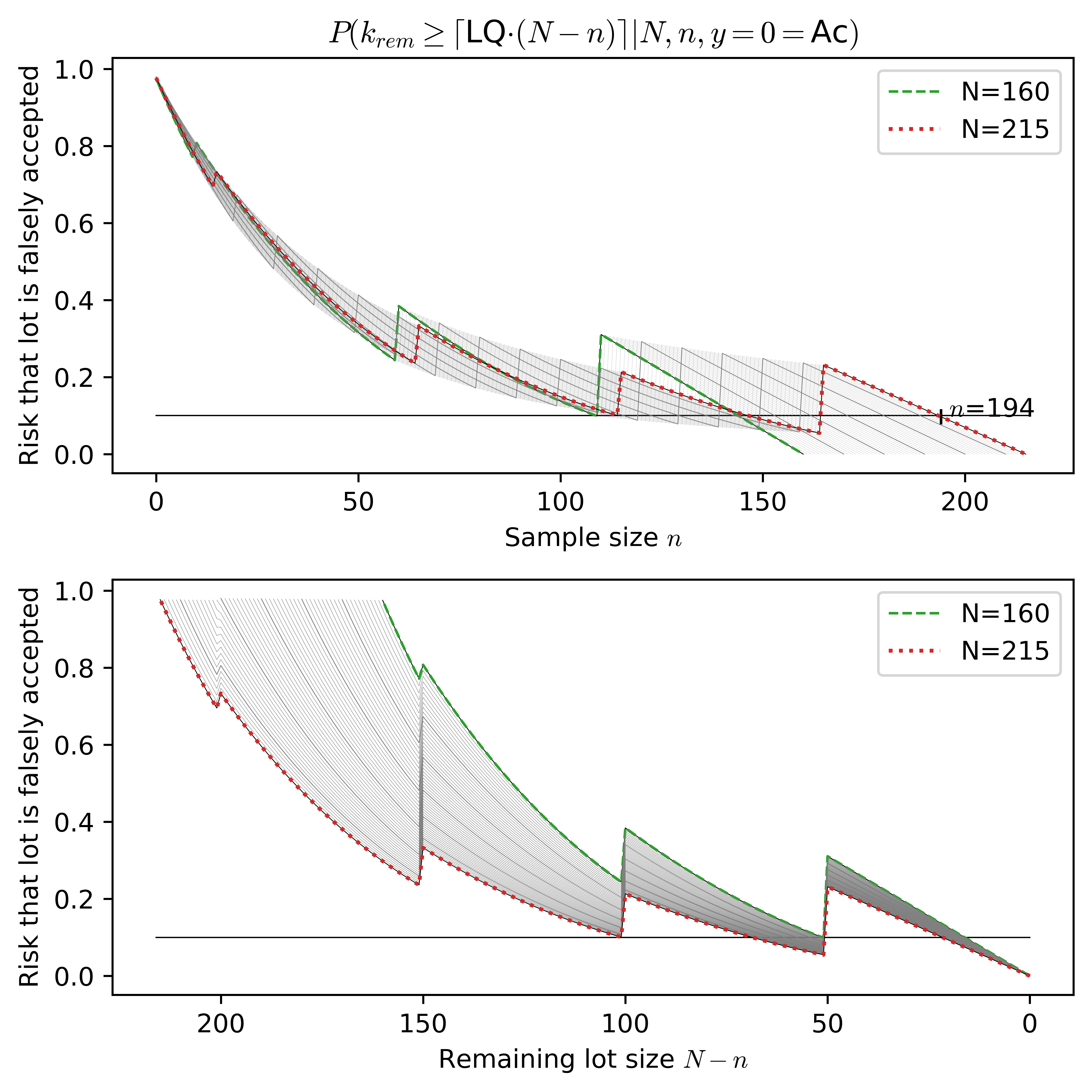}
     \caption{Specific consumer's risk as a function of the sample size (top) and the remaining lot size (bottom) for LQ = 2~\% and lot sizes $N\in\{160,161,...,215\}$}.
     \label{fig:compare_exact_lot_sample_size}
 \end{figure}

To construct more efficient tabulated plans for destructive sampling that assess remaining lots, let us explain the substantial difference in sample sizes between plans for single lot sizes and for ranges of lot sizes.
For this, we visualized the risk~\eqref{EqBayesRisk} for the range $\{160,161,...,215\}$ of lot sizes over the sample size $n$ in the top of Figure~\ref{fig:compare_exact_lot_sample_size}.  
Considering only one lot size, e.g.\ the green curve where $N=160$, we see that the risk decreases in $n$ as long as $ \lceil \text{LQ}\cdot (N-n)\rceil$ remains unchanged.
When the unacceptable number of  nonconforming items in the remaining lot $ \lceil \text{LQ}\cdot (N-n)\rceil$  decreases (due to an increasing $n$), the risk~\eqref{EqBayesRisk} can increase (marked by the 'jumps' in Figure~\ref{fig:compare_exact_lot_sample_size}).
The minimal sample size for lot size $N$ is the smallest one where the corresponding curve in Figure~\ref{fig:compare_exact_lot_sample_size} is below the risk of 10~\%.
For $N=160$ this is the case at $n=109$.
To find a minimal sample size for a range of lot sizes, the risk curves need to be below the prespecified 10~\% for all the lot sizes simultaneously. 
For the given range, this is $n=194$; a sample size that leaves no remaining lot for most of the lot sizes in this range.
The reason for such an inefficient sample size is that for different lot sizes the risk increases at different sample sizes and exceeds the limit when the sample size is larger than minimal.

We conclude in the interim, that sampling plans limiting the risk \eqref{EqBayesRisk} by 10~\% for all lots sizes in a range are inefficient if the risks exceed the limit for sample sizes slightly larger than optimal. This can be observed when the lot is small and the sample size is relatively large.

\subsubsection*{Tabulated plans for destructive sampling}
To overcome the problem of inefficient sampling plans for ranges of lot sizes and enable to provide affordable, tabulated sampling plans for destructive sampling that assess remaining lots, we suggest a new representation: 
State the sampling plan as the pair $[N-n,\text{Ac}]$ instead of  $(n,\text{Ac})$, if beneficial for a range of lot sizes. The risks as a function of the remaining lot size increase only at multiples of 1/LQ (see bottom of Figure~\ref{fig:compare_exact_lot_sample_size}), which allows us to determine more efficient sampling plans for ranges of small lot sizes.
For example, a remaining lot size $N-n=51$ limits the risk \eqref{EqBayesRisk} by 10~\% simultaneously for the range $N\in\{160,161,...,215\}$.
We denote these plans by square brackets $[N-n,\text{Ac}]$ to distinguish them from conventional plans $(n,\text{Ac})$.

Table~\ref{TabPlans} column 2 and Figure~\ref{FigOptimSP} in blue display sampling plans that limit the risk~\eqref{EqBayesRisk} by 10~\% for LQ = 2~\%.
\begin{table*} 
	\centering
		\begin{tabular}{rc|rc}
         Lot size& Sampling plan &         Lot size  & Sampling plan\\
			       &for  LQ = 2~\%        & &for LQ = 20~\%\\\hline
            0-49 & no plan & 0-16& no plan\\
			50-98 & [5, 0] & 17-21 & ~[6, 0]\\
			99-159 & [10, 0] & 22-26 & [11, 0]\\
			160-215 & [51, 0]  & 27-31 & [16, 0]\\
			216-266 & [101, 0] & 32-35 & [21, 0]\\
			267-316 & [151, 0] & 36-40 & [26, 0]\\[4pt]
            317-500&(154, 0)&41-50&(12, 0)\\
			501-1200 & (132, 0) & 51-90 & (12, 0)
		\end{tabular}
	\caption{Tabulated sampling plans for destructive sampling by attribute that limit the specific consumer's risk \eqref{EqBayesRisk} by 10~\% for a uniform prior and acceptance numbers $\text{Ac}=0$.}
	\label{TabPlans}
\end{table*}
The ranges of lot sizes are chosen here such that the difference between neighboring optimal sample sizes (black) is small in each range. 
As a result, the difference between the sample sizes for the tabulated plans (blue) and the optimal sample sizes (black) is less than 1/LQ.
 An analogue sampling plan for LQ = 20~\% is displayed in column 4 of Table~\ref{TabPlans} to indicate the tabulation of sampling plans for different values of LQ.
 For large lot sizes, sampling plans of the form ($n$, Ac) are more suitable for tabulation, because they allow to combine large ranges of lot sizes with only small reductions of efficiency.

To summarize, sampling plans provided in the square bracket representation which limit the risk \eqref{EqBayesRisk}, as those in Table~\ref{TabPlans},
 \begin{itemize}
     \item are suitable for destructive testing,
     \item are presented concisely,
     \item provide an almost minimal required sample size in particular for small lots, and
     \item do not require software for their application.
 \end{itemize}
 Concise and efficient tables similar to Table~\ref{TabPlans} can also be generated when the prior distribution is informative or when variable sampling is conducted. 
 Additionally, they may be designed to limit some producer's risk or minimize some costs.
 
 Therefore, we recommend the new square bracket representation as a starting point for standardizing destructive sampling plans that assess remaining lots.

\subsubsection*{Prior sensitivity analysis}
The sampling plans suggested in Sections~\ref{SecBayes} and \ref{SecTabs} limit the risk~\eqref{EqBayesRisk} by 10~\% based on the uniform prior for the number of nonconforming items in the lot before sampling.
The sampling plans in square bracket representation (c.f.\ Section~\ref{SecTabs}) limit the risk~\eqref{EqBayesRisk} also when applied for larger lots;
that is, when sampling zero nonconforming items from a lot of size $N'\geq N$ and such that the remaining lot still has the same size $N-n$.
 In this section, we examine how sensitive the developed plans are with respect to changes in the prior distribution.
 In particular, we fix the sampling plans [$N-n,$ Ac] from Table \ref{TabPlans} and show how the range of lot sizes needs to change for the plans to still limit the risk~\eqref{EqBayesRisk} by 10~\%
 when a beta-binomial prior $P(k_{whole})$ with parameters $N,a,$ and $b$ is used:
 \begin{align*}
     P(k_{whole})&=\text{Beta-binomial}(k_{whole};N,a,b)\\
     &=\binom{N}{k_{whole}}\frac{B(k_{whole}+a,N-k_{whole}+b)}{B(a,b)},
 \end{align*}
 where $B$ is the beta function.
 That is, sensitivity is considered within the class of conjugate priors.

\begin{figure}
    \centering
    \begin{subfigure}{0.5\linewidth}\includegraphics[scale=.63]{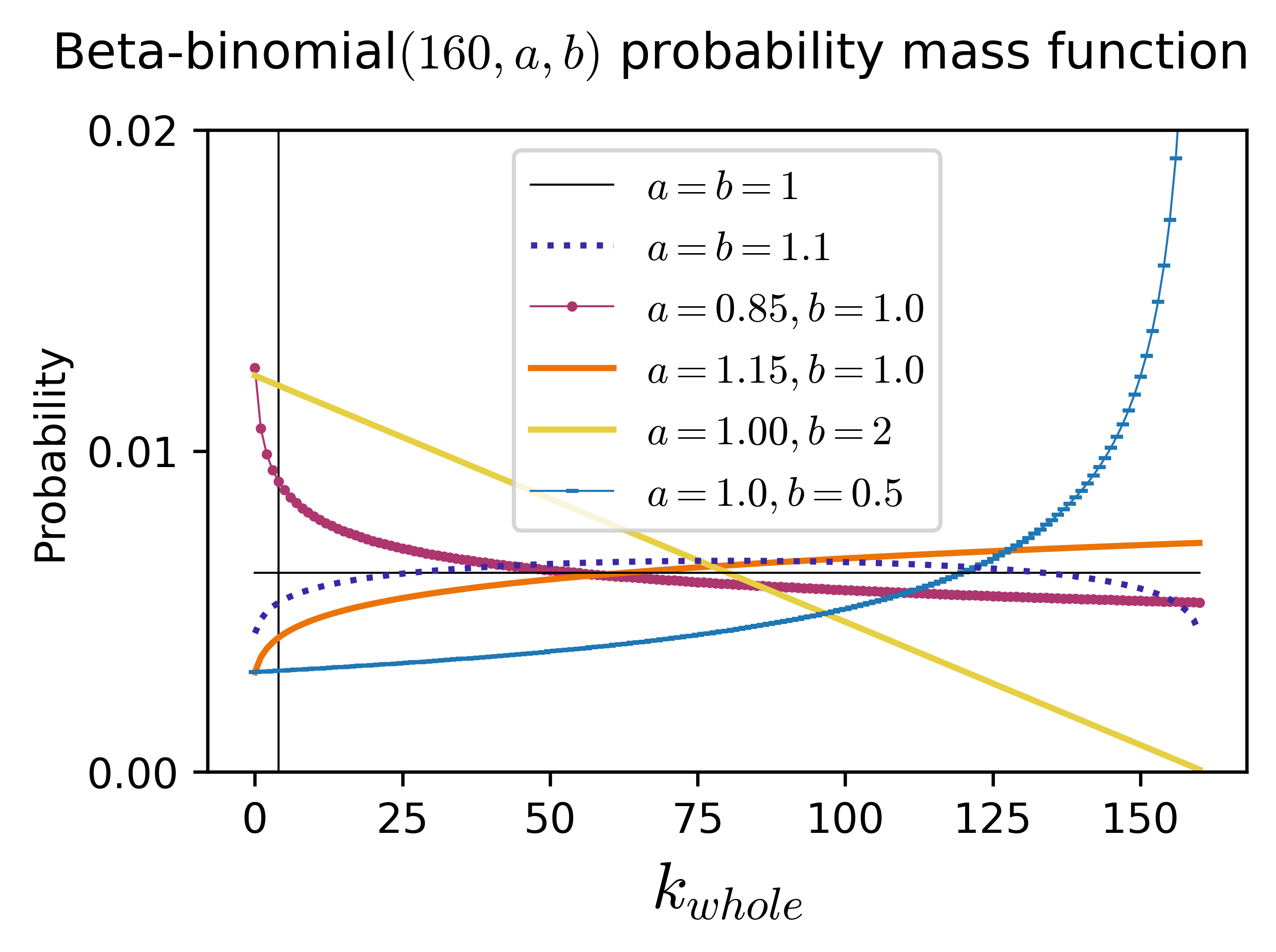}
\end{subfigure}\vline~\begin{subfigure}{0.5\linewidth}
    \includegraphics[scale=.63]{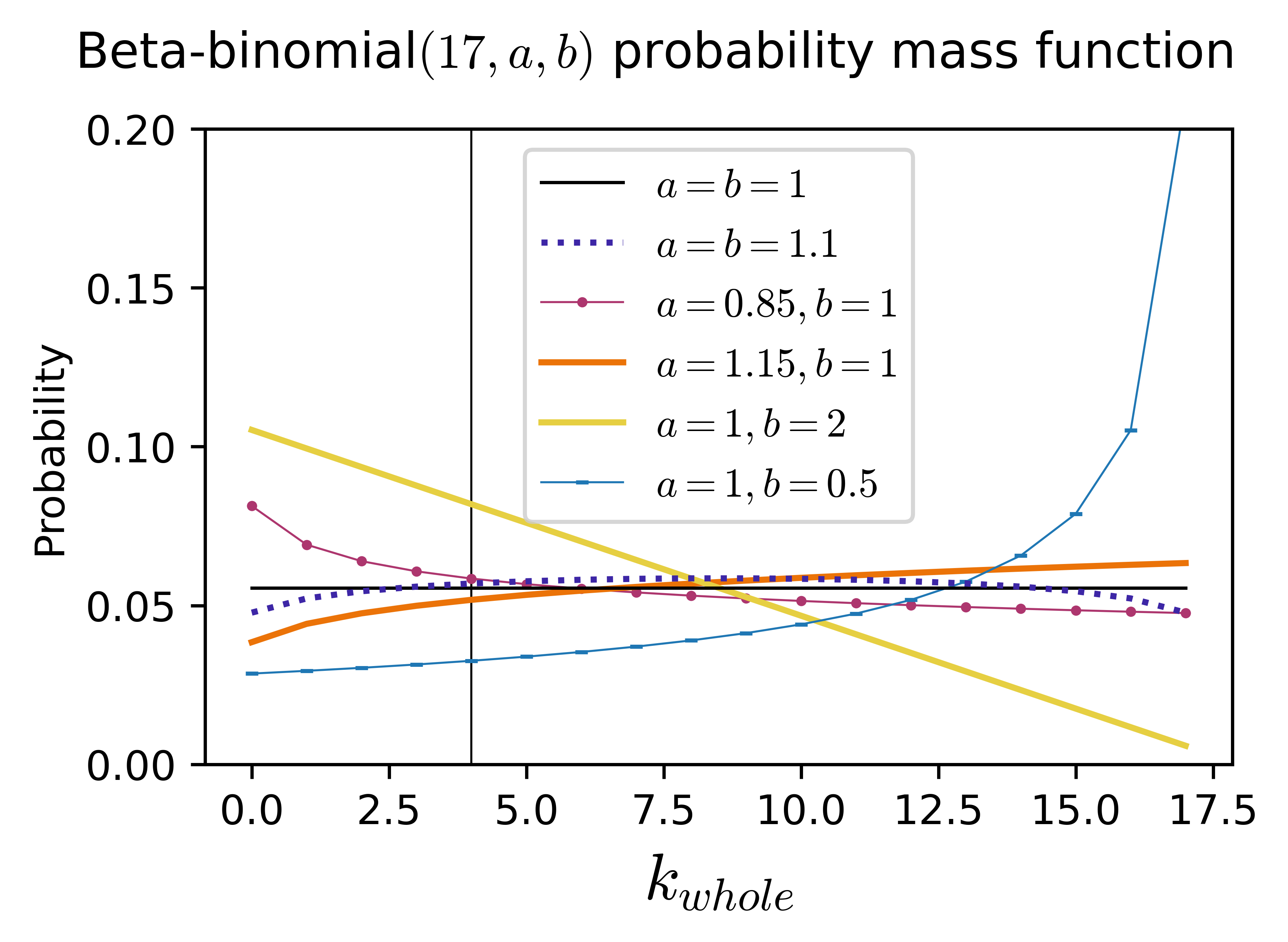}
\end{subfigure}
    \caption{The beta-binomial distribution for a set of different parameters is shown for a lot of size 160 (left) and 17 (right). The vertical line is positioned at the best unsatisfactory quality before sampling $\lceil N\, \text{LQ}\rceil$ (left: LQ = 2~\%, right: LQ = 20~\%). }
    \label{fig:beta-binomialPrior}
\end{figure}

Figure~\ref{fig:beta-binomialPrior} shows prior distributions for different parameters
to give an intuition on how much the prior changes with its parameters.
We see, for example, that a change from $a=1$ to $a=0.85$ or $a=1.15$ roughly doubles or halves the weights of the prior at perfect quality, respectively.
Or, if we change the parameters from $a=b=1$ to $a=b=1.1$, the prior weights of perfect quality are reduced by a quarter already.
The hyperparameters $a$ and $b$ can be interpreted as \textit{a priori} data, where $a-1$ is the number of prior nonconforming items and $b-1$ is the number of prior conforming items (c.f. \cite[p. 34]{GelmanAndrew2014Bda}).

\begin{figure}
\centering
\begin{minipage}[t]{.5\linewidth}
\begin{subfigure}{\linewidth}
    \includegraphics[scale=.67]{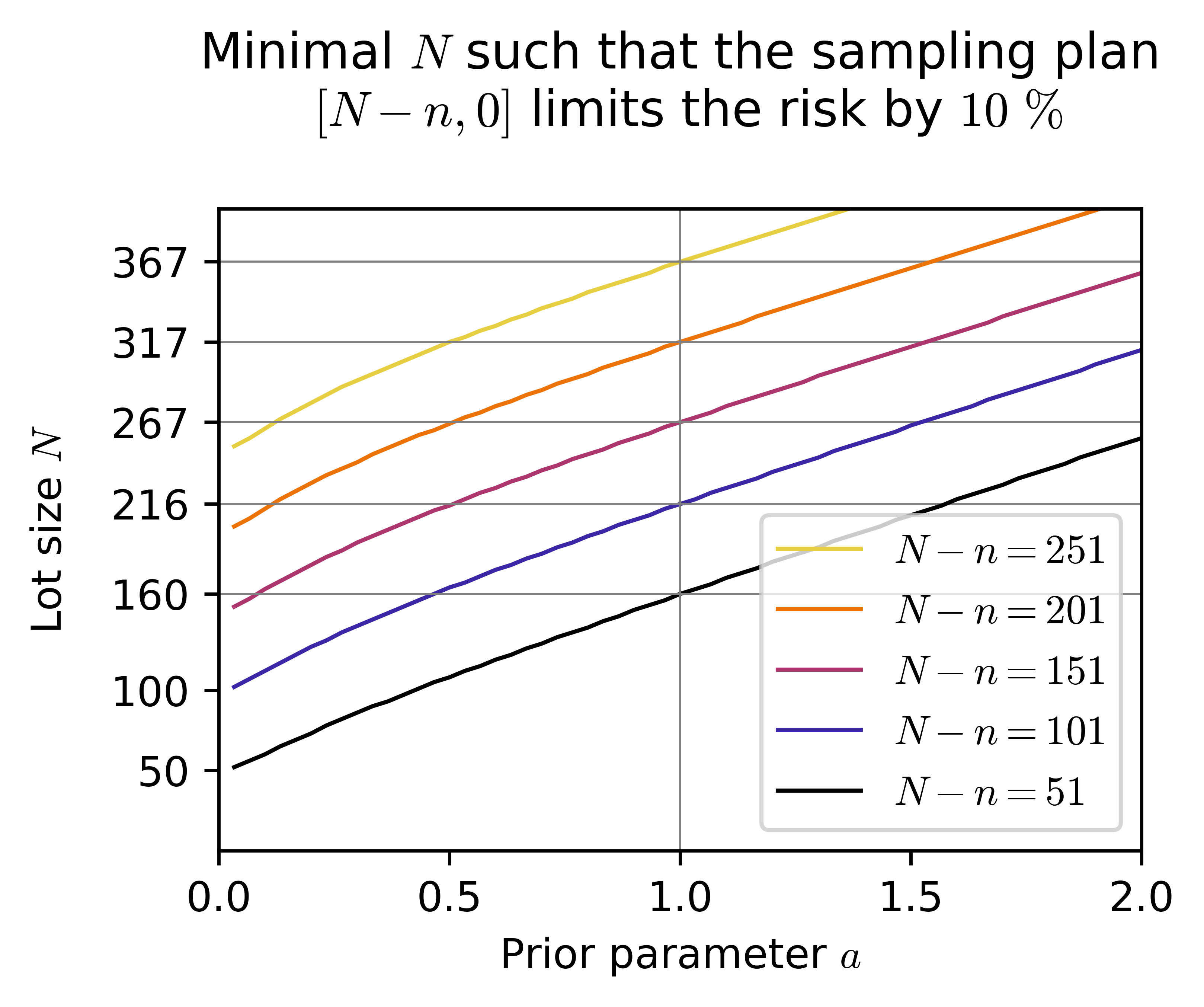}
    \caption{LQ = 2~\%, $b=1$.}
    \label{fig:LQ2bfix}
\end{subfigure}\\
\begin{subfigure}{\linewidth}
    \includegraphics[scale=.67]{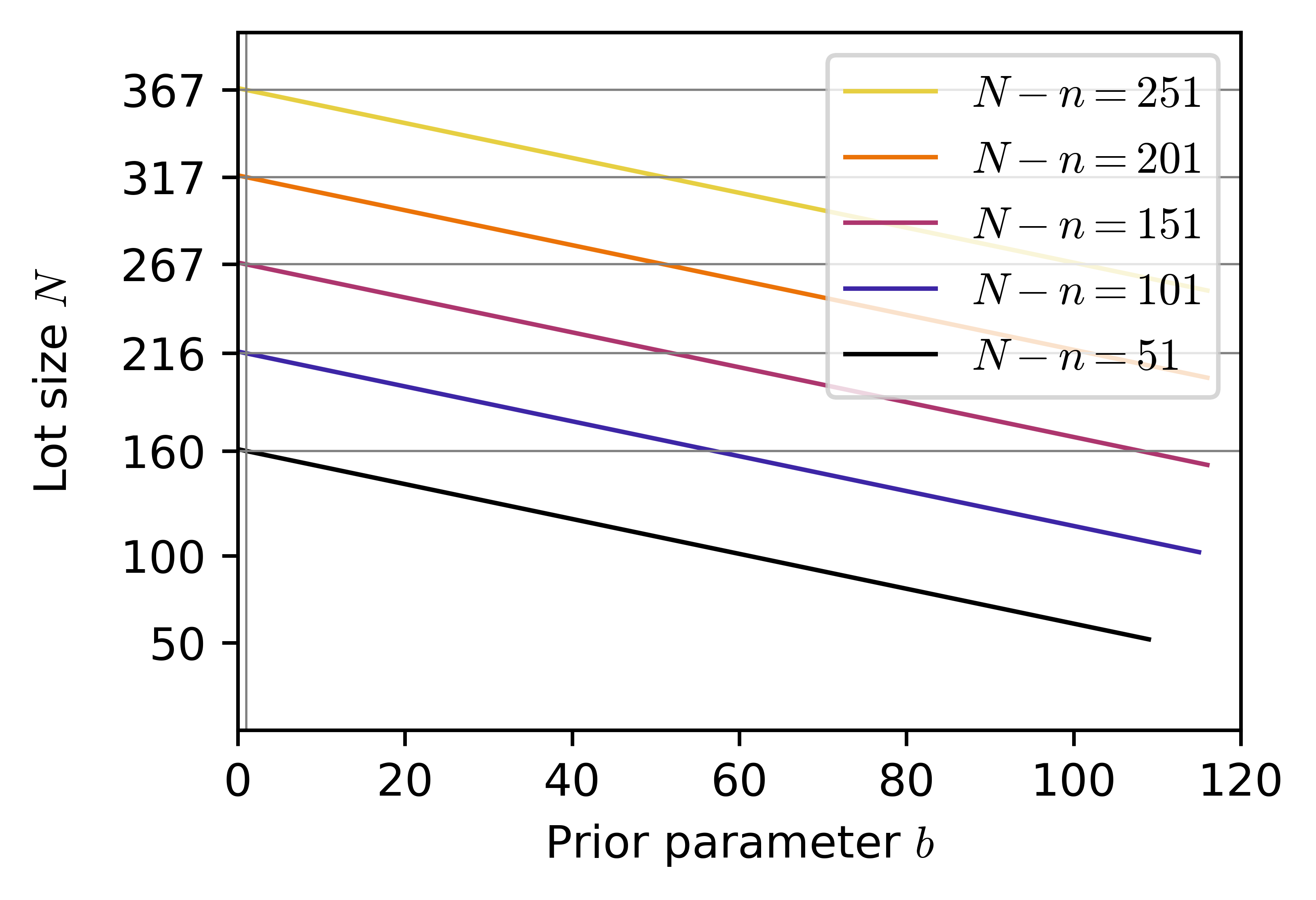}
    \caption{LQ = 2~\%, $a=1$.}
    \label{fig:LQ2afix}
\end{subfigure}\\
\begin{subfigure}{\linewidth}
    \includegraphics[scale=.67]{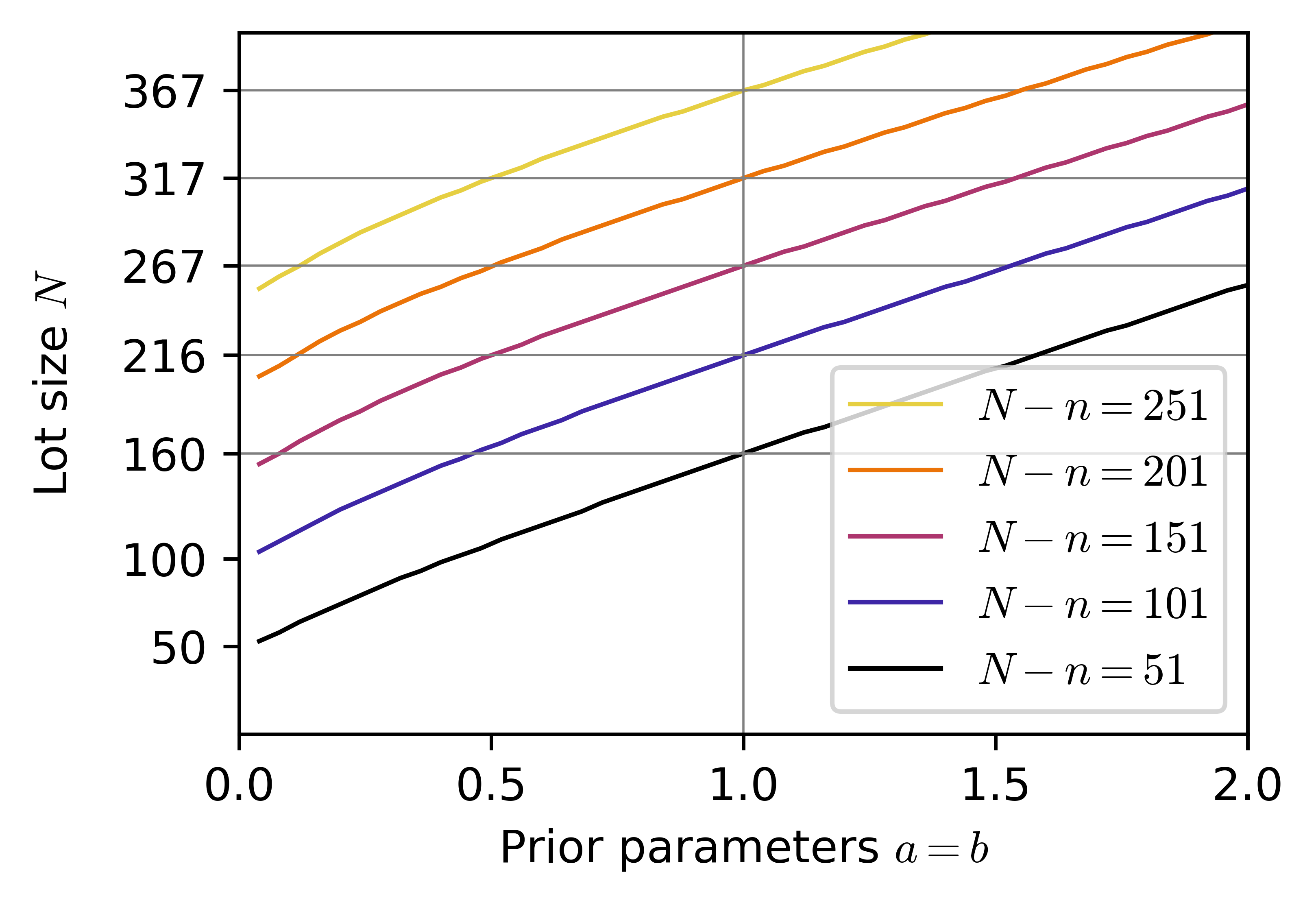}
    \caption{LQ = 2~\%, $a=b$.}
    \label{fig:LQ2a=b}
\end{subfigure}
\end{minipage}\vline~~~~\begin{minipage}[t]{.5\linewidth}
\begin{subfigure}{\linewidth}
    \includegraphics[scale=.67]{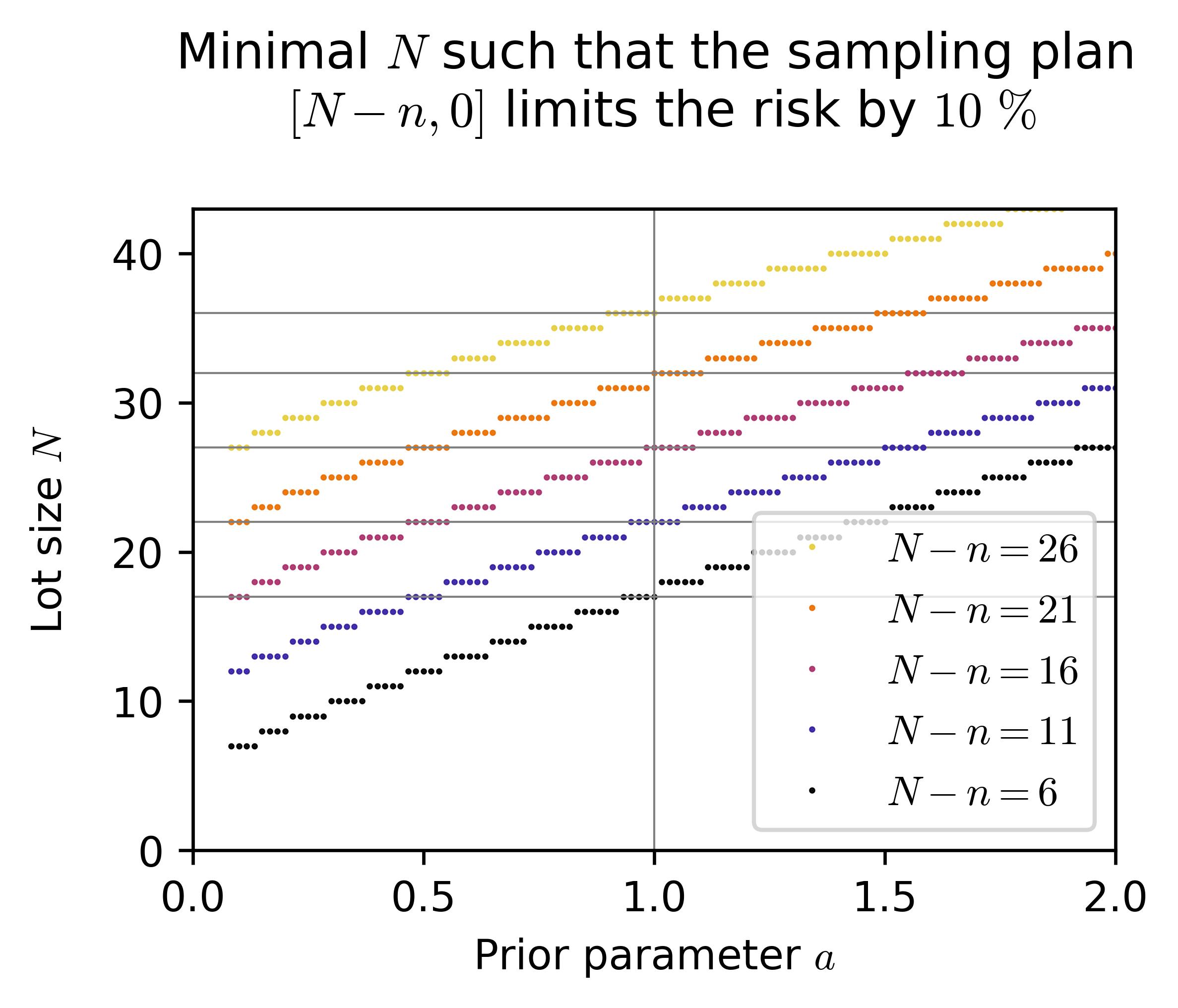}
    \caption{LQ = 20~\%, $b=1$.}
    \label{fig:LQ20bfix}
\end{subfigure}\\
\begin{subfigure}{\linewidth}
    \includegraphics[scale=.67]{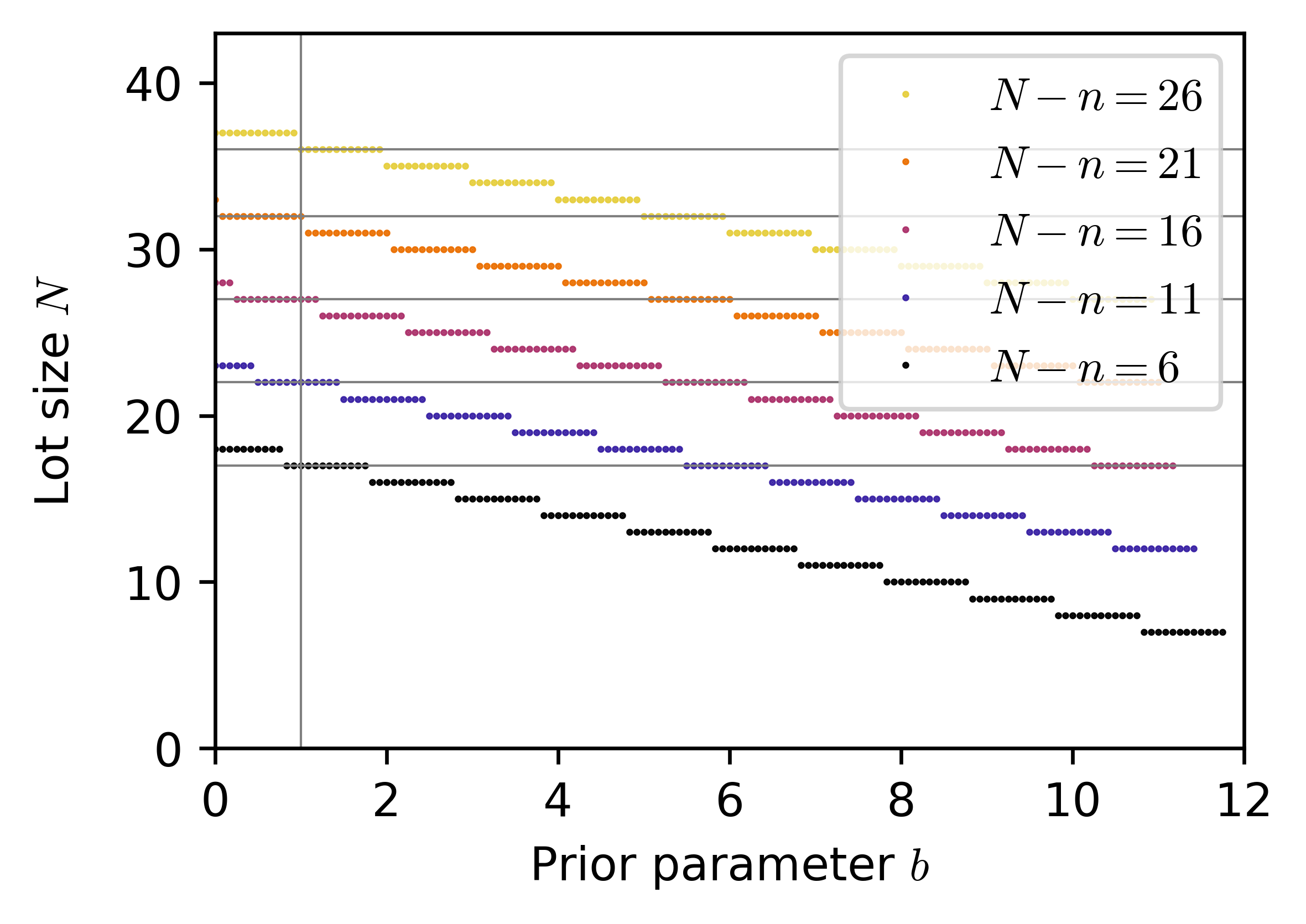}
    \caption{LQ = 20~\%, $a=1$.}
    \label{fig:LQ20afix}
\end{subfigure}\\
\begin{subfigure}{\linewidth}
    \includegraphics[scale=.67]{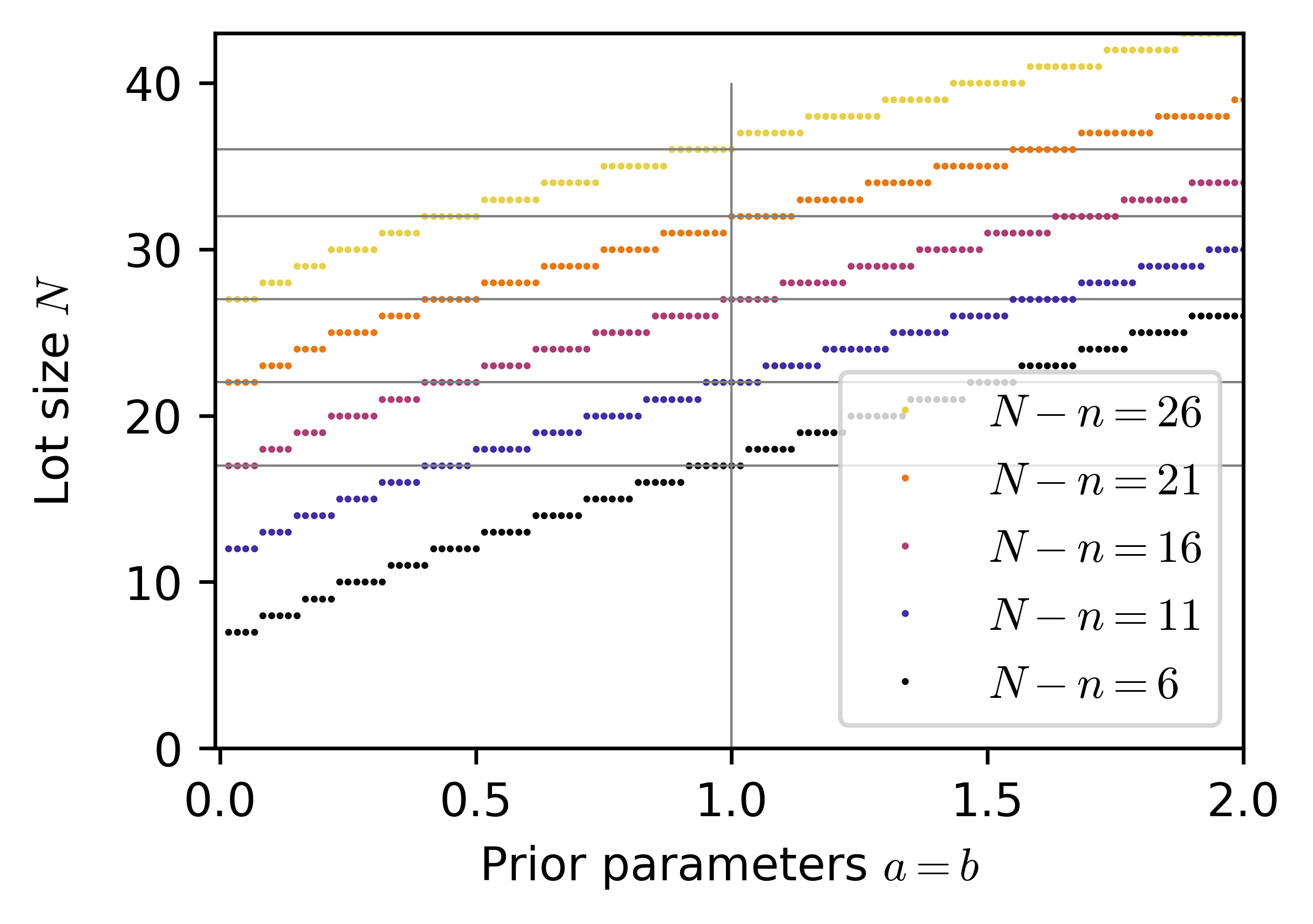}
    \caption{LQ = 20~\%, $a=b$.}
    \label{fig:LQ20a=b}
\end{subfigure}
\end{minipage}
    \caption{The minimal required lot sizes for the prespecified sampling plans $[N-n, 0]$ to limit the risk~\eqref{EqBayesRisk} by 10~\% are plotted over the parameters $a$ or $b$ of the beta-binomial prior distribution and for two different LQ values.}
    \label{fig:SensAnalysis}
\end{figure}

We discuss now the sensitivity of the sampling plans for LQ $=2~\%$ and display the results in Figures~\ref{fig:LQ2bfix}-\ref{fig:LQ2a=b}.
The results for LQ = 20~\% are shown in Figures~\ref{fig:LQ20bfix}-\ref{fig:LQ20a=b}  but are not discussed as they are qualitatively similar to LQ~=~2~\%.
We conjecture that the results below are transferable also to other LQs.

If we fix the prior parameter $b$ and increase $a$, the prior has more weight on unacceptable quality (compare Figure~\ref{fig:beta-binomialPrior}).
In this case, we need to increase the lot sizes in the table such that the risk \eqref{EqBayesRisk} is still limited to 10~\%.
If we fix $b$ and decrease $a$, the prior has more weight on acceptable quality.
In this case, we can add smaller lot sizes to the ranges in the table leaving  the risk~\eqref{EqBayesRisk} limited to 10~\%.
The dependency is displayed in Figure~\ref{fig:LQ2bfix}.
It shows the minimal $N$ over the prior parameter $a$ such that the sampling plans [51,~0], [101,~0], [151,~0], [201,~0], and [251,~0] limit the risk~\eqref{EqBayesRisk}.
Remember, that there is no maximal $N$ for  sampling plans of the square braked form. 
We see that the parameter $a$ needs to decrease by about 0.5 to apply a more efficient  plan (i.e.\ with smaller sample size) from the table to the same range of lot sizes.
The slope of the minimal required lot size $N$ over the prior parameter $a$ is about 100. That is, if the parameter $a$ changes by 0.1, the minimal required lot sizes $N$ change by about 10.

A similar, but reversed dependency can be observed when fixing the prior parameter $a$ and varying $b$, as displayed in Figure~\ref{fig:LQ2afix}.
We see that we need to increase the parameter $b$ by about 50 to 60 to apply a more efficient sampling plan from the table.
The slope of the minimal required lot size $N$ over the prior parameter $b$ is about $-1$.

If we increase the prior parameters $a$ and $b$ equally, the prior's variance decreases and the distribution has less weight at the boundaries.
The dependency of the prior on the minimal required lot size limiting the risk~\eqref{EqBayesRisk} by 10~\%  is displayed in Figure~\ref{fig:LQ2a=b} and shows a very similar behavior to when $b$ is fixed and $a$ varies (compare Figure~\ref{fig:LQ2bfix}).

In conclusion, the ranges of lot sizes are quite robust to changes in the prior distribution. That is,
small changes in the prior distribution require only small changes in the range of lot sizes such that the sampling plans still limit the specific consumer's risk \eqref{EqBayesRisk}.
The ranges of lot sizes are more sensitive to the parameter $a$ than to the parameter $b$ of the beta-binomial prior, especially when LQ is small.
Varying the parameter $b$ has a one-to-one impact on the required sample size.
The impact of increasing $a$ or decreasing $b$ may be high for the smallest lots of the range, 
because a small increase in the specific consumer's risk may lead to exceeding its limit.
In such a case one can either increase the consumer's risk limit slightly, or design and use a plan with larger sample size.
When $a$ decreases or $b$ increases, the sampling plans protect the consumers even better.
In such a case the sampling plans do not need to be changed but their sample sizes can be reduced optionally.

\section{Conclusions}
Often acceptance sampling is performed for attributive observations which are destructive. The aim is then to assess the quality of the lot without that sample.

This research shows that the frequentist consumer's risk  to accept a remaining lot with unsatisfactory quality cannot be described by the hypergeometric distribution.
In addition, the percentage of nonconforming items in the whole lot needs to be lower than limiting quality in order for an accepted remaining lot to have at least limiting quality.
Consequently, the sampling plans in the international standard ISO 2859-2 for attribute sampling in isolated lots are ill-suited for destructive sampling and we recommend future editions to explicitly state this limitation. 
The authors are not aware of other standards for destructive attribute sampling that assess remaining lots.

The Bayesian approach is suitable to design sampling plans for destructive sampling, for example to limit the specific consumer’s risk of insufficient quality in the remaining lot.
The sampling plans provided in the standard ISO 2859-2 do not limit this Bayesian risk by 10~\%. 
In fact, this risk can exceed 44~\%.
No other standard is available yet that provides sampling plans limiting the specific consumer's risk.  
This research works towards closing this gap and suggests to efficiently tabulate plans that limit the specific consumer’s risk by fixing the remaining lot size $N-n$ instead of fixing the sample size $n$ for ranges of small lot sizes. 
That is, we recommend to represent sampling plans for small lot sizes as pairs [$N-n$, Ac] instead of the conventional pairs ($n$, Ac), with Ac being the usual acceptance number. 
We provide such tabulated sampling plans for a limiting quality of 2~\% and 20~\% based on a so-called non-informative prior distribution and a risk limit of 10~\%.
However, this new representation in square brackets is also applicable to informative priors or to variable sampling.
For the tabulated sampling plans,  the applicable range of lot sizes varies linearly when varying the parameters of the conjugate prior distribution.
For the smallest (but not for the larger) lots of a range, small changes of the prior distribution towards unacceptable qualities can require
to slightly increase the consumer's risk limit, or to use plans with larger sample sizes.
The tabulated sampling plans are robust to moving prior weight towards acceptable quality.

In conclusion, this research is the first to provide tabulated sampling plans for destructive sampling by attributes  that assess remaining lots. The resulting tables are simple to use, still efficient and thus provides a valuable contribution to the standardization of destructive sampling.

\section*{CRediT author statement} 
\textbf{Hugalf Bernburg:} conceptualization (equal); formal analysis (lead); investigation (lead); methodology (lead); software (lead); visualization (equal); writing – original draft (supporting); writing – review and editing (equal)\\
\textbf{Katy Klauenberg:} conceptualization (equal); formal analysis (supporting); investigation (supporting); methodology (supporting); project administration; software (supporting); supervision; validation; visualization (equal); writing – original draft (lead); writing – review and editing (equal)

\begin{acks}[Acknowledgments]
Corresponding author is H. Bernburg: hugalf.bernburg@ptb.de.

The authors are grateful to their colleague and supervisor Clemens Elster, who passed away during the course of this research. With his strong network and wide expertise in statistics and its application in metrology, he contributed valuable advice and discussions to this research.
\end{acks}

\begin{funding}
Support from the German Research Foundation through the project with number 547236699 is gratefully acknowledged.
\end{funding}

\bibliographystyle{imsart-nameyear}
\bibliography{literatur}

\begin{thebibliography}{27}

\bibitem[\protect\citeauthoryear{{ISO/TC 69/SC 1}}{2006}]{ISO3534-2}
\begin{bmisc}[author]
\bauthor{\bsnm{{ISO/TC 69/SC 1}}}
(\byear{2006}).
\btitle{ISO 3534-2 Statistics -- Vocabulary and symbols, Part 2: Applied
  statistics}.
\bhowpublished{{International Organization for Standardization (ISO)}}.
\end{bmisc}
\endbibitem

\bibitem[\protect\citeauthoryear{{OIML TC 3/SC 4}}{2017}]{OIMLG20}
\begin{barticle}[author]
\bauthor{\bsnm{{OIML TC 3/SC 4}}}
(\byear{2017}).
\btitle{Surveillance of utility meters in service on the basis of sampling
  inspections}.
\bjournal{OIML G 20}.
\end{barticle}
\endbibitem

\bibitem[\protect\citeauthoryear{{ISO/TC 69/SC 5}}{2005}]{ISO2859-3}
\begin{bmisc}[author]
\bauthor{\bsnm{{ISO/TC 69/SC 5}}}
(\byear{2005}).
\btitle{ISO 2859-3 Sampling procedures for inspection by attributes -- Part 3:
  Skip-lot sampling procedures}.
\bhowpublished{{International Organization for Standardization (ISO)}}.
\end{bmisc}
\endbibitem

\bibitem[\protect\citeauthoryear{{ISO/TC 69/SC 5}}{2017}]{ISO28590}
\begin{bmisc}[author]
\bauthor{\bsnm{{ISO/TC 69/SC 5}}}
(\byear{2017}).
\btitle{ISO 28590 Sampling procedures for inspection by attributes --
  Introduction to the ISO 2859 series of standards for sampling for inspection
  by attributes}.
\bhowpublished{{International Organization for Standardization (ISO)}}.
\end{bmisc}
\endbibitem

\bibitem[\protect\citeauthoryear{{ISO/TC 69/SC 5}}{2020}]{ISO2859-2}
\begin{bmisc}[author]
\bauthor{\bsnm{{ISO/TC 69/SC 5}}}
(\byear{2020}).
\btitle{ISO 2859-2 Sampling procedures for inspection by attributes -- Part 2:
  Sampling plans indexed by limiting quality (LQ) for isolated lot inspection}.
\bhowpublished{{International Organization for Standardization (ISO)}}.
\end{bmisc}
\endbibitem

\bibitem[\protect\citeauthoryear{{ISO/TC 690}}{2017}]{ISO28594}
\begin{bmisc}[author]
\bauthor{\bsnm{{ISO/TC 690}}}
(\byear{2017}).
\btitle{ISO 28594 Combined accept-zero sampling systems and process control
  procedures for product acceptance}.
\bhowpublished{{International Organization for Standardization (ISO)}}.
\bnote{(en)}.
\end{bmisc}
\endbibitem

\bibitem[\protect\citeauthoryear{Anderson and Anderson}{1996}]{Anderson96}
\begin{barticle}[author]
\bauthor{\bsnm{Anderson},~\bfnm{Craig~A.}\binits{C.~A.}} \AND
  \bauthor{\bsnm{Anderson},~\bfnm{Kathryn~B.}\binits{K.~B.}}
(\byear{1996}).
\btitle{Violent crime rate studies in philosophical context: {A} destructive
  testing approach to heat and southern culture of violence effects}.
\bjournal{Journal of personality and social psychology}
\bvolume{70}
\bpages{740}.
\end{barticle}
\endbibitem

\bibitem[\protect\citeauthoryear{Avellaneda, Melo and
  Cruz}{2024}]{Avellaneda24}
\begin{barticle}[author]
\bauthor{\bsnm{Avellaneda},~\bfnm{C.~A.}\binits{C.~A.}},
  \bauthor{\bsnm{Melo},~\bfnm{O.~O.}\binits{O.~O.}} \AND
  \bauthor{\bsnm{Cruz},~\bfnm{N.~A.}\binits{N.~A.}}
(\byear{2024}).
\btitle{Analysis of longitudinal data with destructive sampling using linear
  mixed models}.
\bjournal{arXiv preprint arXiv:2411.16153}.
\end{barticle}
\endbibitem

\bibitem[\protect\citeauthoryear{Bernardo}{2003}]{bernardo2003bayesStatis}
\begin{binbook}[author]
\bauthor{\bsnm{Bernardo},~\bfnm{Jos{\'e}~M.}\binits{J.~M.}}
(\byear{2003}).
\btitle{Bayesian Statistics}
In \bbooktitle{Probability and {S}tatistics}
\bvolume{2}.
\bpublisher{{E}ncyclopedia of Life Support Systems ({EOLSS})}.
\end{binbook}
\endbibitem

\bibitem[\protect\citeauthoryear{BIPM et~al.}{2012}]{JCGM106}
\begin{bmisc}[author]
\bauthor{\bsnm{BIPM}}, \bauthor{\bsnm{IEC}}, \bauthor{\bsnm{IFCC}},
  \bauthor{\bsnm{ILAC}}, \bauthor{\bsnm{ISO}}, \bauthor{\bsnm{IUPAC}},
  \bauthor{\bsnm{IUPAP}} \AND \bauthor{\bsnm{OIML}}
(\byear{2012}).
\btitle{Evaluation of measurement data --- {T}he role of measurement
  uncertainty in conformity assessment}.
\bhowpublished{Joint Committee for Guides in Metrology, JCGM 106:2012}.
\bdoi{10.59161/JCGM106-2012}
\end{bmisc}
\endbibitem

\bibitem[\protect\citeauthoryear{Delgadillo, Bremer and
  Hoffman}{2007}]{Delgadillo07}
\begin{barticle}[author]
\bauthor{\bsnm{Delgadillo},~\bfnm{Francisco}\binits{F.}},
  \bauthor{\bsnm{Bremer},~\bfnm{Ron}\binits{R.}} \AND
  \bauthor{\bsnm{Hoffman},~\bfnm{James~J.}\binits{J.~J.}}
(\byear{2007}).
\btitle{A destructive sampling method designed for outsourcing situations
  involving high quality production processes}.
\bjournal{Quality \& quantity}
\bvolume{41}
\bpages{513--529}.
\end{barticle}
\endbibitem

\bibitem[\protect\citeauthoryear{Dodge and Romig}{1954}]{Dodge1944Sampling}
\begin{bbook}[author]
\bauthor{\bsnm{Dodge},~\bfnm{Harold~F.}\binits{H.~F.}} \AND
  \bauthor{\bsnm{Romig},~\bfnm{Harry~G.}\binits{H.~G.}}
(\byear{1954}).
\btitle{Sampling inspection tables : single and double sampling},
\bedition{6. print} ed.
\bseries{Wiley publications in statistics}.
\bpublisher{Wiley publications in statistics}, \baddress{New York u.a.}
\end{bbook}
\endbibitem

\bibitem[\protect\citeauthoryear{Dyer and Pierce}{1993}]{Dyer93}
\begin{barticle}[author]
\bauthor{\bsnm{Dyer},~\bfnm{Danny}\binits{D.}} \AND
  \bauthor{\bsnm{Pierce},~\bfnm{Rebecca~L.}\binits{R.~L.}}
(\byear{1993}).
\btitle{On the choice of the prior distribution in hypergeometric sampling}.
\bjournal{Communications in Statistics-Theory and Methods}
\bvolume{22}
\bpages{2125--2146}.
\end{barticle}
\endbibitem

\bibitem[\protect\citeauthoryear{Gelman}{2014}]{GelmanAndrew2014Bda}
\begin{bbook}[author]
\bauthor{\bsnm{Gelman},~\bfnm{Andrew}\binits{A.}}
(\byear{2014}).
\btitle{Bayesian data analysis},
\bedition{3} ed.
\bseries{Chapman \& Hall/CRC texts in statistical science series}.
\bpublisher{CRC Press}, \baddress{Boca Raton u.a.}
\end{bbook}
\endbibitem

\bibitem[\protect\citeauthoryear{Gorman and Bower}{2002}]{Gorman02}
\begin{barticle}[author]
\bauthor{\bsnm{Gorman},~\bfnm{Douglas}\binits{D.}} \AND
  \bauthor{\bsnm{Bower},~\bfnm{Keith~M.}\binits{K.~M.}}
(\byear{2002}).
\btitle{Measurement Systems Analysis and Destructive Testing}.
\bjournal{Six Sigma Forum Magazine}
\bvolume{1}
\bpages{16-19}.
\end{barticle}
\endbibitem

\bibitem[\protect\citeauthoryear{Hald}{1960}]{Hald60}
\begin{barticle}[author]
\bauthor{\bsnm{Hald},~\bfnm{Anders}\binits{A.}}
(\byear{1960}).
\btitle{The compound hypergeometric distribution and a system of single
  sampling inspection plans based on prior distributions and costs}.
\bjournal{Technometrics}
\bvolume{2}
\bpages{275--340}.
\end{barticle}
\endbibitem

\bibitem[\protect\citeauthoryear{Herman and
  Robbins}{2013}]{herman2013Hypergeometric}
\begin{barticle}[author]
\bauthor{\bsnm{Herman},~\bfnm{Rod~A.}\binits{R.~A.}} \AND
  \bauthor{\bsnm{Robbins},~\bfnm{Kelly~R.}\binits{K.~R.}}
(\byear{2013}).
\btitle{Use of hypergeometric distribution for estimating adventitious presence
  of GM traits in small seed lots may be misleading}.
\bjournal{Seed Science Research}
\bvolume{23}
\bpages{211--212}.
\end{barticle}
\endbibitem

\bibitem[\protect\citeauthoryear{Hsu}{1977}]{Hsu77}
\begin{barticle}[author]
\bauthor{\bsnm{Hsu},~\bfnm{John I.~S.}\binits{J.~I.~S.}}
(\byear{1977}).
\btitle{A cost model for skip-lot destructive sampling}.
\bjournal{IEEE Transactions on Reliability}
\bvolume{26}
\bpages{70--72}.
\end{barticle}
\endbibitem

\bibitem[\protect\citeauthoryear{Insua et~al.}{2020}]{Insua20}
\begin{barticle}[author]
\bauthor{\bsnm{Insua},~\bfnm{David~Rios}\binits{D.~R.}},
  \bauthor{\bsnm{Ruggeri},~\bfnm{Fabrizio}\binits{F.}},
  \bauthor{\bsnm{Soyer},~\bfnm{Refik}\binits{R.}} \AND
  \bauthor{\bsnm{Wilson},~\bfnm{Simon}\binits{S.}}
(\byear{2020}).
\btitle{Advances in Bayesian decision making in reliability}.
\bjournal{European Journal of Operational Research}
\bvolume{282}
\bpages{1--18}.
\end{barticle}
\endbibitem

\bibitem[\protect\citeauthoryear{Johnson, Kotz and Kemp}{1992}]{Johnson92}
\begin{barticle}[author]
\bauthor{\bsnm{Johnson},~\bfnm{N.~L.}\binits{N.~L.}},
  \bauthor{\bsnm{Kotz},~\bfnm{S.}\binits{S.}} \AND
  \bauthor{\bsnm{Kemp},~\bfnm{A.~W.}\binits{A.~W.}}
(\byear{1992}).
\btitle{Discrete univariate distributions}.
\bjournal{Wiley series in probability and mathematical statistics (Probability
  and mathematical statistics)}.
\end{barticle}
\endbibitem

\bibitem[\protect\citeauthoryear{Patel}{1989}]{patel1989prediction}
\begin{barticle}[author]
\bauthor{\bsnm{Patel},~\bfnm{J.~K.}\binits{J.~K.}}
(\byear{1989}).
\btitle{Prediction intervals-a review}.
\bjournal{Communications in Statistics-Theory and Methods}
\bvolume{18}
\bpages{2393--2465}.
\end{barticle}
\endbibitem

\bibitem[\protect\citeauthoryear{Phelps}{1982}]{Phelps82}
\begin{barticle}[author]
\bauthor{\bsnm{Phelps},~\bfnm{R.~I.}\binits{R.~I.}}
(\byear{1982}).
\btitle{Skip-lot destructive sampling with Bayesian inference}.
\bjournal{IEEE Transactions on Reliability}
\bvolume{31}
\bpages{191--193}.
\end{barticle}
\endbibitem

\bibitem[\protect\citeauthoryear{Uhlig et~al.}{2024}]{Uhlig24}
\begin{barticle}[author]
\bauthor{\bsnm{Uhlig},~\bfnm{Steffen}\binits{S.}},
  \bauthor{\bsnm{Colson},~\bfnm{Bertrand}\binits{B.}},
  \bauthor{\bsnm{Kissling},~\bfnm{Roger}\binits{R.}},
  \bauthor{\bsnm{Ellis},~\bfnm{Sam}\binits{S.}},
  \bauthor{\bsnm{Hicks},~\bfnm{Mike}\binits{M.}},
  \bauthor{\bsnm{Vandenbemden},~\bfnm{John}\binits{J.}},
  \bauthor{\bsnm{Pennecchi},~\bfnm{Francesca}\binits{F.}},
  \bauthor{\bsnm{G{\"o}b},~\bfnm{Rainer}\binits{R.}} \AND
  \bauthor{\bsnm{Gowik},~\bfnm{Petra}\binits{P.}}
(\byear{2024}).
\btitle{Acceptance Sampling Plans Based on Conformance Probability --
  Inspection of Lots and Processes by Attributes}.
\bjournal{{preprints.org}}.
\end{barticle}
\endbibitem

\bibitem[\protect\citeauthoryear{Wallenius}{1967}]{wallenius1967sampling}
\begin{barticle}[author]
\bauthor{\bsnm{Wallenius},~\bfnm{K.~T.}\binits{K.~T.}}
(\byear{1967}).
\btitle{Sampling for confidence}.
\bjournal{Journal of the American Statistical Association}
\bvolume{62}
\bpages{540--547}.
\end{barticle}
\endbibitem

\bibitem[\protect\citeauthoryear{WELMEC}{2024}]{WelmecGuide8.10}
\begin{bmisc}[author]
\bauthor{\bsnm{WELMEC}}
(\byear{2024}).
\btitle{Measuring Instruments Directive (2014/32/EU): Guide for generating
  sampling plans for statistical verification according to Annex F and F1 of
  MID 2014/32/EU}.
\bhowpublished{{WELMEC European cooperation in legal metrology: Working Group
  8}}.
\end{bmisc}
\endbibitem

\bibitem[\protect\citeauthoryear{Wilson}{1984}]{Wilson84}
\begin{bbook}[author]
\bauthor{\bsnm{Wilson},~\bfnm{Forrest}\binits{F.}}
(\byear{1984}).
\btitle{Building materials evaluation handbook}.
\bpublisher{Van Nostrand Reinhold Company Inc.}
\end{bbook}
\endbibitem

\bibitem[\protect\citeauthoryear{Wilson and Farrow}{2021}]{Wilson21}
\begin{barticle}[author]
\bauthor{\bsnm{Wilson},~\bfnm{Kevin~J.}\binits{K.~J.}} \AND
  \bauthor{\bsnm{Farrow},~\bfnm{Malcolm}\binits{M.}}
(\byear{2021}).
\btitle{Assurance for sample size determination in reliability demonstration
  testing}.
\bjournal{Technometrics}
\bvolume{63}
\bpages{523--535}.
\end{barticle}
\endbibitem

\end{thebibliography}
\end{document}